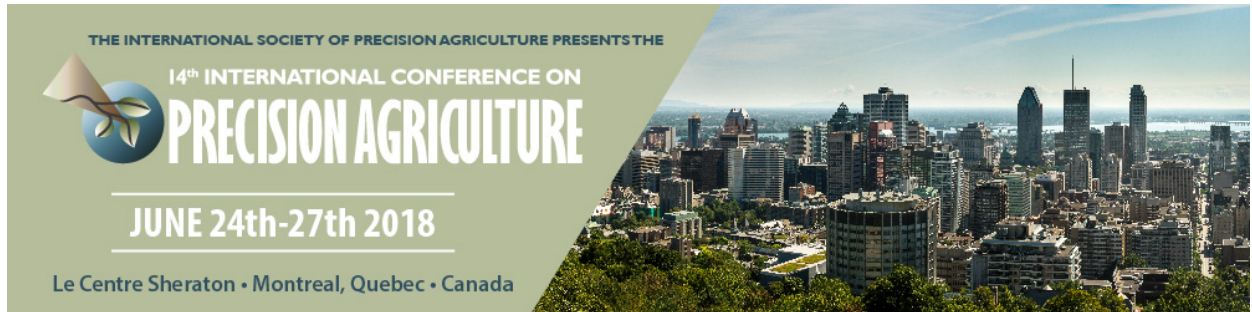

# Exploring Wireless Sensor Network Technology In Sustainable Okra Garden: A Comparative Analysis Of Okra Grown In Different Fertilizer Treatments

**Lamar Burton, Yemeserach Mekonnen, Arif Sarwat, Shekhar Bhansali, Krish Jayachandran**

*Florida International University, College of Electrical and Computer Engineering*
*Miami, Florida, USA*
lburt004@fiu.edu
ymeko001@fiu.edu
asarwat@fiu.edu
sbhansa@fiu.edu
jayachan@fiu.edu

**Abstract**:

*The goal of this project was to explore commercial agricultural and irrigation sensor kits and to discern if the commercial wireless sensor network (WSN) is a viable tool for providing accurate real-time farm data at the nexus of food energy and water. The smart garden consists of two different varieties of Abelmoschus esculentus (okra) planted in raised beds, each grown under two different fertilizer treatments. Soil watermark sensors were programed to evaluate soil moisture and dictate irrigation events up to four times a day, while soil temperature and photosynthetic solar radiation sensors also recorded data every six hours. Solar panels harvested energy to power water pump and sensors. The objectives of the experiments were to evaluate and compare plant and soil parameters of the two okra varieties grown under two different fertilizer treatments. The plant parameters evaluated and compared were basal diameter, plant height, fruit production, and fruit size. Soil parameters measured were soil moisture, soil temperature, and soil nitrate concentration. The commercial sensors were evaluated on efficiency, accuracy, ease of use and overall practicality. Clemson spineless produced larger okra plants with the highest plant parameter values, followed by Emerald okra. However, they both averaged nearly the same yield and length of okra fruit. Nature's Care fertilizer leached more in beds containing Clemson spineless, while Garden-tone leached more in beds containing Emerald okra. When the WSN is installed properly, the system's great performance undoubtedly aides the farmer by providing real time field data. However, a properly installed apparatus does not promise a stable system. There are numerous challenges and limitations of which can diminish the performance quality of the WSN, those being battery power, data transmission, and data storage. Data storage is also an issue depending on the amount of data collected, rate of data collection, and size of storage unit. These issues can hinder the decision making for precision farmers.*

**Introduction**

According to the Food and Agriculture Organization (FAO), the world's population is expected to increase to 9.6 Billion people by the year 2050, and as the population continues to grow exponentially, unfavorable climatic conditions frequently threatens food security. In efforts to combat world demand for food in a rapidly changing and unpredictable climate, farmers are exploring smart agriculture to help satisfy desired production outputs while conserving precious resources. Smart agriculture or smart farming, offers new techniques using wireless sensor network (WSN) technology to continuously monitor plant and soil parameters and more accurately control fertigation events. Fertigation is the application of fertilizer and soil amendment with irrigation water, and is also a commonly used method in agriculture because of its time saving feature.

Modern methods of farming such as the "touch-and-feel" method for determining soil moisture can now be thing of the past as sensor technology presents a novel approach to precision agriculture. Precision agriculture, as defined by the United States Department of Agriculture (USDA), is an information and technology based management system that is site specific and uses one or more of the following sources of data: soils, crops, nutrients, pests, moisture, or yield, for optimum profitability, sustainability, and protection of the environment.

However, the use of WSN is still newly explored and majority of farmers have not fully adopted the method and instead uses traditional methods for determining the need of fertilizer and water. These poor methods produce waste and utilize excess resources such as energy, water, and fertilizers that damage our ecosystems. For this reason, more research has been geared towards technology which can enhance food and environmental security, minimize production costs, and optimize profits for farmers.

To date, WSN's have been deployed in environmental and agricultural fields for numerous applications such as to manage water resources, manage product storage facilities, determine optimal harvest time, characterize crop growth, and predict fertilizer requirements. Among the many, only these seven are referenced in this paper: a wireless sensor network was deployed in [1] to monitor water content, temperature and salinity of soil at a cabbage farm located in a semi-arid region of Spain. This experiment tested the range, robustness, and reliability of the wireless sensor network system by analyzing the network's performance and energy consumption. In [2] temperatures at various positions in a feed warehouse were monitored using a wireless sensor network. Common issues with feed barns is that changes in temperature excites the growth of certain bacteria and fungus, which can thereby cause disease to human and farm animals. The energy limitations of wireless sensor networks were the focus of [3]. Another WSN was deployed in [4] to monitor a greenhouse environment. Another greenhouse monitoring system was deployed in [5] and it's communication was based on Zigbee. A system designed to capture farmland information was deployed in [6] to study the transmission signal in different growth periods on a wheat field. A WSN was applied to a water saving irrigation system in [7].

The above citations prove the existence and research done in efforts to advance the study of WSN in precision agriculture. In fact, these sensing platforms are available for purchase today, however their expensiveness makes them unattractive to farmers who desire their use. Most farmers can only afford to deploy one sensing platform per acre, which is inefficient because farm terrain varies spatially. What is present or lacking in one area of terrain may not be suitable for another area within that acre. If sensors can be made inexpensively with the ability to measure high density and high resolution, then this data obtained can be sent cheaply to a cloud source where data can be analyzed using computational dynamic models to not only predict the future leaching behavior but also use inverse modeling to pinpoint sources of their leaching behavior. There are endless opportunities for the application of these sensors if these sensor systems can be made cheaply enough so they can be usable worldwide, given that agriculture regions are scattered making terrains and environmental conditions vary significantly. Therefore, creating methods to accurately and rapidly quantify spatial variation for crop growth and soil quality should be further investigated.

*Nitrogen Cycle / Nitrogen Runoff*

Nitrogen is the major limiting macronutrient for plant growth, it is considered a popular soil nutrient to both commercial farmers and home gardeners. On the farm, animal urine, feces, and fertilizers contribute to majority of nitrogen found in soils. The most common form of nitrogen that is plant-available in soils is nitrate ($NO_3^-$), however the element nitrogen can also be present in soils as ammonium ($NH_4^+$), nitrite ($NO_2^-$), and or organic matter. The high mobilizing property of the nitrate ions allows it to easily maneuver through soils, therefore leaching and running off into waterways. If the soil biome consists of certain anaerobic bacteria, nitrate can be converted into the gaseous form of nitrogen, which is released into the atmosphere. Currently, standard laboratory techniques use chemical and optic sensors for determining total N, ammonium, nitrate, and nitrite.

As stated earlier, nitrogen in excess during irrigation and rainfall is subject to runoff and leaching into surface and groundwater. An accumulation of nitrogen has effects that are hazardous to humans and the marine ecosystem. High concentrations of nitrogen in the form of nitrate in underground water sources poses a threat to humans consuming it, and when nitrates runoff into surface water, nitrate buildup can lead to a polluted state toxic to marine life known as eutrophication. In [8] and [9], researchers in China addresses special issues of eutrophication such as the increased activity of algal blooms and fish kills as well as investigate nitrate leaching under intensive vegetable garden production patterns. The conclusions indicate the need for innovative technology that can better monitor soil quality for the protection of the environment.

**Materials and methods**

The Abelmoschus esculentus (okra) raised garden beds were constructed on March 25, 2017 at the Florida International University Engineering Center (FIU EC). The farm consists of three 4 feet x 22 feet garden beds filled with 12 cubic yards of loam soil purchased from Home Depot in Miami, Florida. Cow manure was also purchased and mixed in each bed as amendments at the time of construction. A 10 feet x 3 feet solar panel was used to harvest energy. Blue Sky Solar Boost MPPT charge controller was purchased from Alternative Energy Store Inc. Boxborough, Massachusetts and delivered 15 amps to successfully charge a 24-volt system used to power the 24-volt DC water pump also purchased from Alternative Energy Store Inc. The Shurflow water pump transferred fluids from the storage tank to the sprayer with an inlet pressure of 2.06 bar (30 psi) and an output flow rate of 4 gallons per minute, providing test beds with water through DIG drip irrigation, purchased from Home Depot Miami, Florida. The drip hose contained emitters spaced six inches apart releasing water at a rate of 25 gallons per hour.

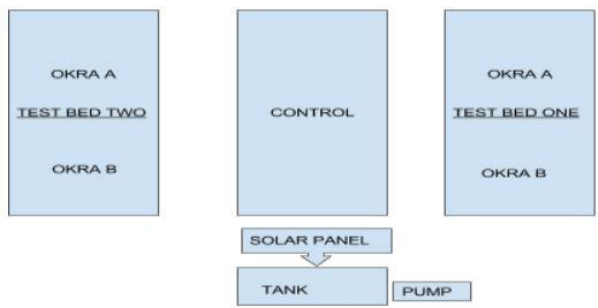

Figure: Okra farm

Commercial WSN kit was purchased from Libelium in Zaragoza, Spain and were assembled at the okra garden on March 26, 2017. The set up included a Waspmote Agriculture Sensor Board Pro, which served as a micro-controller, a Waspmote ZB Pro SMA 5dBi radio for transmitting data packets, 6600mAh rechargeable battery, 7.4-volt solar panel, watermark soil moisture sensor, solar radiation / ultraviolet sensor, soil temperature Pt-1000 sensor, and a Meshlium ZigBee PRO access point.

Soil moisture and temperature sensor values identified accurate thresholds for soil moisture content and can also be used to dictate irrigation events. The watermark sensor is a resistive type sensor consisting of two anti-corrosive electrodes embedded in a granular matrix beneath a gypsum wafer. The values are directly proportional to the soil water tension and returns the frequency output of the sensor's adaptation in a hertz within the range of 50 - 10000 hz. The wetter the soil the higher the frequency, the dryer the soil the lower the frequency output. Watermark sensor was also assembled to arduino microcontroller and values received dictated commands given to water pump for irrigating the okra beds. The Pt-1000 is also a resistive based sensor which returns the temperature values in degree Celsius. The solar radiation sensor provides at its output a voltage proportional to the intensity of the light in the ultraviolet range of the spectrum.

All data obtained from sensors are forwarded to Meshlium access point and stored directly to the hard drive or sent to a cloud service. Meshlium is a Linux router which works as the gateway of the waspmote sensor network. Inserting a sim onto the waspmote sim slot allows for data and commands to be transmitted to cellular devices.

The ZigBee radio used to transmit data frames to the meshlium operates at 2.54 Ghz using a transmission power of only 50 mW and a line of sight 5dBi dipole antenna to cover a range of 7000 meters. An illustration of the WSN is below.

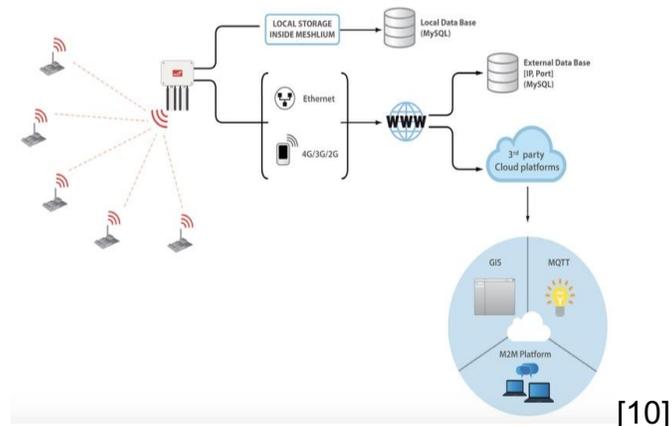 [10]

Figure: Sensor network

Sensor based technology for precision agriculture were used to evaluate the effects of two different organic fertilizers on nitrate leaching and plant growth parameters of two okra varieties. The organic fertilizers, Nature's Care (3-4-2) by Miracle Grow, and Espoma's Garden-tone (3-4-4), were compared against a controlled group of 21 okra plants grown without fertilizers. Nature's Care is a granular fertilizer with 3% slowly available water insoluble nitrogen (N) derived from feather meal, fish meal, blood meal, wheat middlings, and meat and bone meal. Espoma's Garden-tone is also a granular fertilizer, however it's Total nitrogen contain 0.2% ammoniacal nitrogen, 0.6% other water-soluble nitrogen, and 2.2% water insoluble nitrogen. The water insoluble nitrogen is a slow release nitrogen from hydrolyzed feather meal, pasteurized poultry manure, bone meal and alfalfa meal.

The two Abelmoschus esculentus varieties used in this experiment were: Clemson Spineless, a popular heirloom because of it's high yield and taste, and Quimbombo Emerald, commonly chosen because it bears tender fruit of large size. Clemson spineless were planted on July 22, 2017 and Emerald on July 27, 2017. Experimental plants were grown from seeds, sewn twelve inches apart directly into garden bed soil, while controlled group were grown in 10 L pots of bare loamy soil with no added fertilizers or microbes. All groups were grown under 100% sun with irrigation, weed and pest control done as needed. Plant growth parameters: number of leaves, leaf length, leaf width, shoot length, basal diameter, amount of okra produced and okra size, were measured every week.

Soil samplers were purchased from Soil Moisture Corp. in Santa Barbara, California and used to collect moisture samples from soil. Samplers used are large-volume samplers designed for near surface installation at depths ranging from 15cm to 1.8 meters. The unit consists of a 4.8 cm outside diameter PVC tube with a 200-Kpa porous ceramic cup. Leachate was collected weekly after irrigation and analyzed for nitrate concentration and monthly composite samples were analyzed for nitrate concentration. Horiba commercial nitrate sensor was used to analyze the soil nitrate concentration and data was compared between the two fertilizer treated beds.

**Results**

We compared nitrate concentration in leachate from two different fertilizer beds, plant growth parameters from two different okra varieties, and evaluated soil properties for precision agriculture using wireless sensor network. Results are compiled as figures and tables below, and represents data obtained over a period of three months.

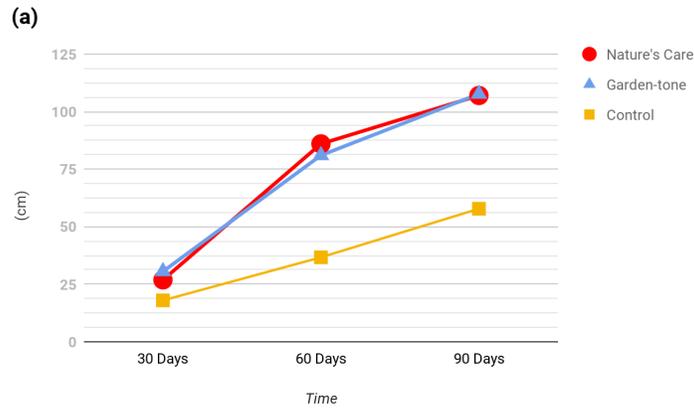

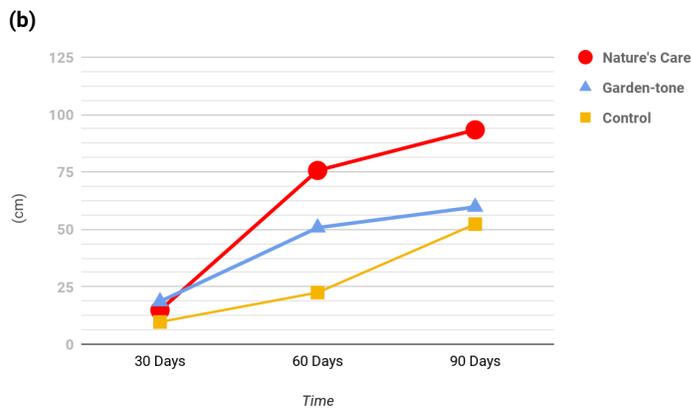

Figure 1: Mean height of Clemson spineless **(a)** and Emerald **(b)** okra plants developed in three months of growth in either Nature's

Figure 1 illustrates the mean height of Clemson spineless and Emerald okra varieties grown for three months in either Nature's Care, Espoma Garden-tone, or no fertilizer. The height of Clemson spineless grown in Nature's Care averaged 26.9 cm at one month, 86 cm at two months, and 107 cm at three months. The height of Clemson spineless grown in Espoma Garden-tone averaged 30.5 cm at one month, 81 cm at two months, and 107.6 cm at three months, while Clemson spineless grown without fertilizer had the poorest average of height with 18 cm at one month, 36.7 cm at two months, and 57.8 cm at the end of three months. The height of Emerald okra grown in Nature's Care averaged 14.8 cm at one month, 75.6 cm at two months, and 93.3 cm at three months. When grown in Espoma Garden-tone, the Emerald averaged 18.6 cm at one month, 50.7 cm at two months, and 59.7 cm at the end of three months, while Emerald grown with no fertilizer averaged 9.7 cm at one month, 22.5 cm at two months, and 52.3 cm at the end of three months. Okra length remained constant between the two different varieties and two different fertilizer treatments, however the okra fruit thickness of Emerald was observed to be quite thinner than that of Clemson spineless.

**Table 1**

### *Clemson Spineless*

| | | | | | |
|---|---|---|---|---|---|
| **Nature's Care** | 9.4 | 107 | 80.8 | 17 | 11.4 |
| **Garden-tone** | 9.25 | 107.6 | 70.5 | 19.6 | 10.1 |
| **Control** | 4.8 | 57.8 | 13.6 | 2 | 7.6 |

### *Emerald*

| | | | | | |
|---|---|---|---|---|---|
| **Nature's Care** | 9.0 | 93.3 | 56.6 | 14.3 | 11 |
| **Garden-tone** | 5.2 | 64.7 | 20.7 | 19.5 | 10.7 |
| **Control** | 5.53 | 52.3 | 16.8 | 2 | 7 |

Table 1: Difference in mean values for basal diameter (cm), height (cm), leaf count per plant, okra count per plant, and fruit size per plant (cm) between 1 and 90 days for Clemson spineless and Emerald okra plants fertilized with Nature's Care, Garden-tone, or no fertilizer treatment.

Difference in plant parameters can be observed in Table: 1 above with Clemson spineless producing the greatest values in plant parameters. Okra basal diameter, height, leaf count, and okra count were notably different between each variety, with Nature's Care fertilizer producing plant parameters with the greater values. Fruit size between both okra varieties, Clemson spineless and Emerald, were nearly identical with a length of 11 cm in both Nature's Care and Espoma's Garden-tone fertilizer treatments. After 90 days, total harvest in pound per week averaged six pounds.

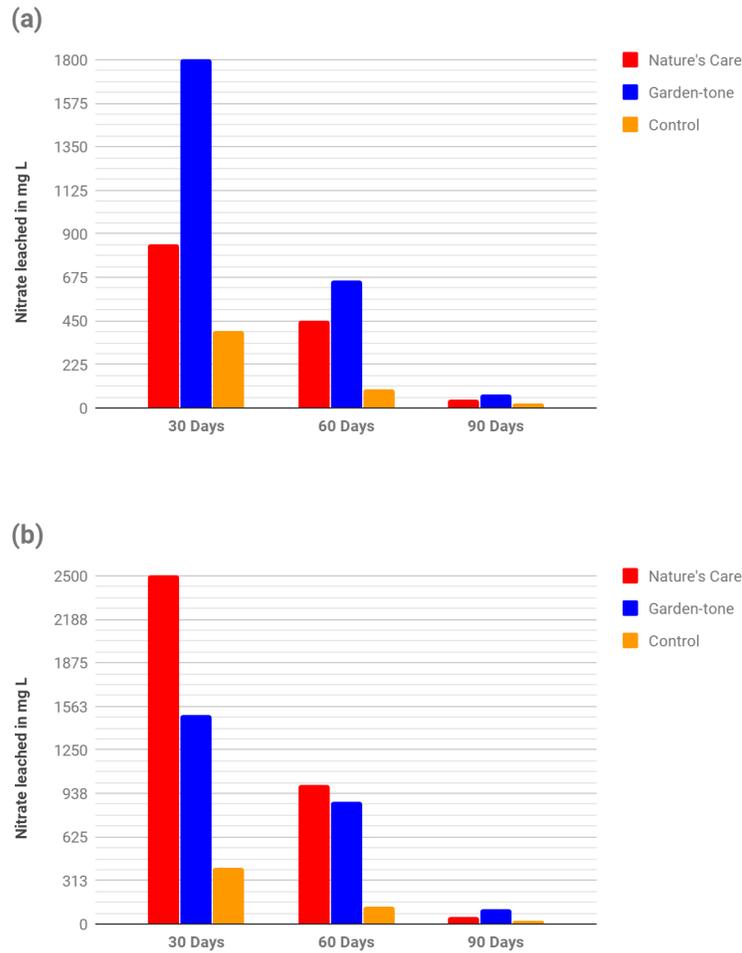

Figure 2: Nitrate leached from soils containing Clemson spineless **(a)** and Emerald **(b)** plants during 3 months of growth with Nature's Care, Espoma Garden-tone, or no fertilizer.

Figure 2 illustrates nitrate leached in soils containing Clemson spineless and Emerald okra grown with under all three treatments. For Clemson spineless okra, Garden-tone was observed to have more leaching with 1800 mg/L at one month, 660 mg/L at month two, and 72 mg/L at month three. Also for Clemson spineless okra, Nature's Care had the second highest values for nitrate leaching with 843 mg/L at month one, 450 mg/L at month two, and 44 mg/L at month three. However, for Emerald okra, Nature's Care had the higher nitrate leachate values for the first two month at 2500 mg/L and 1000 mg/L respectively, followed by 50 mg/L at month three. Garden-tone contained the second highest nitrate leachate values with 1500 mg/L at month one, 875 mg/L at month two, and 108 mg/L at month three.

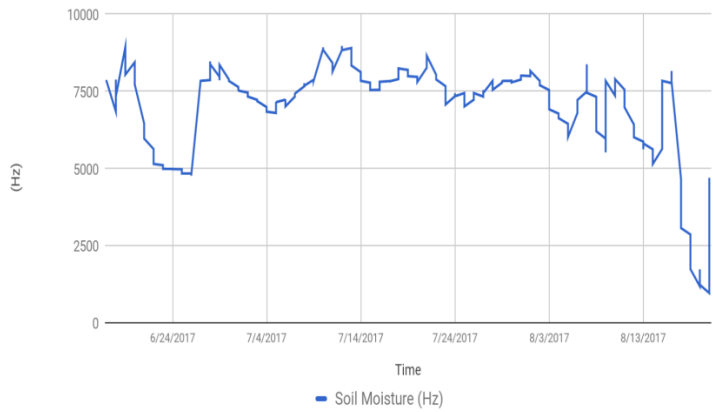

Figure 3: Weekly mean Watermark sensor values for soil moisture (hz) during three months of growth.

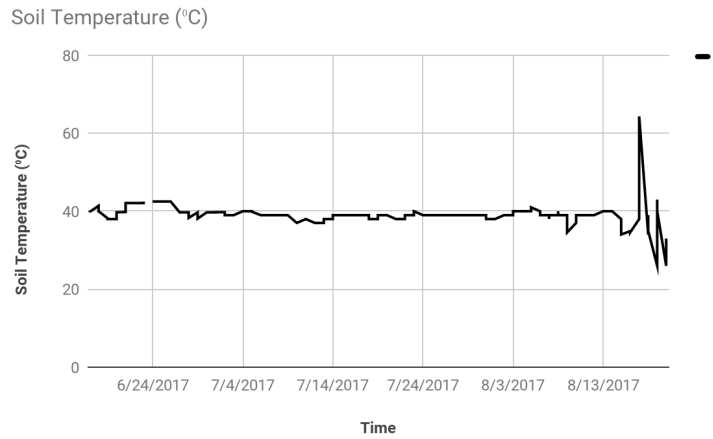

Figure 4: Mean Pt-1000 sensor values for soil temperature (degree celsius) during three months of growth.

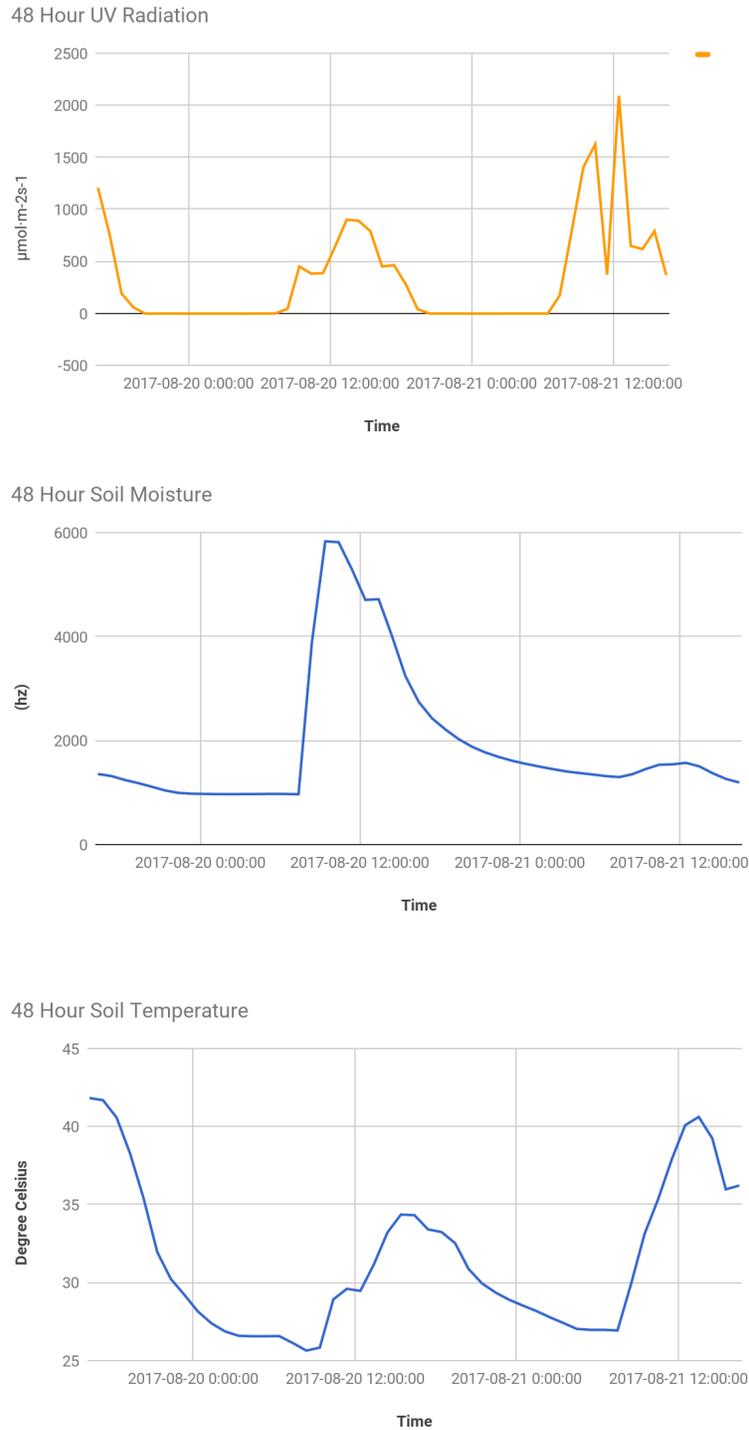

Figure 5: 48 hour ultra violet radiation, soil moisture, and soil temperature values.

We found that the PT1000 soil temperature sensor to be improperly installed at 5 cm depth with part of the sensor exposed, resulting in high temperature readings of 40 degrees Celsius. We suggest that this temperature sensor be buried root deep, as to give a more accurate description of root zone level temperature to correlate with root zone level soil moisture readings. The 48 hour values obtained from watermark, Pt-1000, and UV radiation sensor are strongly correlated. The data suggests a rainfall event occurred in the daytime hours between 0600 hours and 1200 hours.

## Discussion

When the WSN is installed properly, the system's great performance undoubtedly aides the farmer by providing real time field data. However, a properly installed apparatus does not promise a stable system. There are numerous challenges and limitations of which can diminish the performance quality, those being battery power, data transmission, and data storage. Battery power determines the efficiency of the system. When battery level is low, observations were made upon the waspmote's ability to transmit the data packets captured by the sensors. Data storage is also an issue depending on the amount of data collected, rate of data collection, and size of storage unit. These issues can hinder the decision making for precision farmers. Also, the process of determining soil nitrate concentrations is labor intensive and should be further investigated for better methods that will provide real time measurements using wireless sensor technology.

## Future Work

In the future, we plan to use this sensor network setup to investigate soil moisture and temperature readings for optimal lettuce crop production over a longer duration. In addition, we plan to create a more controlled environment with our raised garden beds by employing weed block fabric to the surface of our soil and shade cloth to shield the heat sensitive lettuce crops from the South Florida sun.

## Conclusion

We have explored the use of wireless sensor network technology in a sustainable okra garden, and comparatively analyzed two okra varieties grown under two fertilizer treatments. Clemson spineless produced larger okra plants with the highest plant parameter values, followed by Emerald okra. However, they both averaged nearly the same yield and length of okra fruit. Nature's Care fertilizer leached more in beds containing Clemson spineless, while Garden-tone leached more in beds containing Emerald okra.